\documentstyle[12pt,epsfig]{article}

\newcommand{\bce}{\begin{center}}
\newcommand{\ece}{\end{center}}
\newcommand{\beq}{\begin{equation}}
\newcommand{\eeq}{\end{equation}}
\newcommand{\bea}{\begin{eqnarray}}
\newcommand{\eea}{\end{eqnarray}}
\newcommand{\ba}{\begin{array}}
\newcommand{\ea}{\end{array}}

\newcommand{\doublespace}{
    \renewcommand{\baselinestretch}{1.6}\large\normalsize}

\def\lsim{\mathrel{\rlap{\lower4pt\hbox{\hskip1pt$\sim$}}
    \raise1pt\hbox{$<$}}}         
\def\gsim{\mathrel{\rlap{\lower4pt\hbox{\hskip1pt$\sim$}}
    \raise1pt\hbox{$>$}}}         

\def\Pom{{\bf I\!P}}

\def\PLB{{Phys. Lett.}  B}

\def\ZPC{{Z. Phys.} C}

 \advance\oddsidemargin   by -1in
 \advance\evensidemargin  by -1in
 \advance\topmargin   by -0.5in
\begin{document}
\phantom{.}{\large \bf \hspace{10.2cm} DFTT 33/97
\vspace{.4cm}\\ }

\begin{center}
{\Large \bf Diffraction in Charged Current DIS
\vspace{1.0cm}\\}
{\large \bf M. Bertini$^{a}$, M. Genovese$^{b}$, N.N.~Nikolaev$^{c}$,
B.G.~Zakharov$^{c}$ \vspace{1.0cm}\\}
{\it

$^{a}$ INFN, Sezione di Torino, Via P.Giuria 1, I-10125 Torino, Italy
\bigskip \\
$^{b}$  Istituto Elettrotecnico Nazionale Galileo Ferraris\\
Str. delle Caccie 91\\
10135 Torino, Italy.
\bigskip \\
$^{c}$IKP(Theorie), KFA J{\"u}lich, 5170 J{\"u}lich, Germany
\medskip\\
and L. D. Landau Institute for Theoretical Physics, GSP-1,
117940, \\
ul. Kosygina 2, Moscow V-334, Russia.\vspace{1.0cm}\\ }

\vspace{1cm}

{\Large \bf Abstract} \\
\end{center}
We present the QCD calculation of the diffractive structure function
for  charged current DIS. In particular we analyse the perturbatively
tractable excitation of heavy quarks. We emphasize the peculiarities
of the Regge factorization breaking in excitation of open charm.
\bigskip
\bigskip

\begin{center}
E-mail: kph154@aix.zam.kfa-juelich.de
\end{center}

\vspace{3cm} 

\doublespace
In the last years successful quantitative predictions for
 diffractive electromagnetic Deep
Inelastic Scattering have been obtained in perturbative QCD (pQCD) 
[1 - 10]; for a recent review see \cite{NZDIS97}.

In the next future,
diffraction in charge current (CC) DIS, $e p\rightarrow
\nu p'X$, can shed more light on the pQCD mechanism of Diffractive
DIS. Rapidity gap events in CC DIS have already been observed
at HERA \cite{CCZEUS}
and with amassing more data on CC DIS a detailed comparison
between the experiment and models for diffractive DIS will be
possible \cite{CCDDexp}. Because of the parity non-conservation, in the CC case
one has a larger variety of diffractive SF's
compared to the neutral current  electromagnetic (EM) case.
To the lowest order in pQCD, CC diffractive DIS proceeds by the
Cabibbo-favoured excitation of the $(u\bar{d})$ and $(c\bar{s})$
dijet states. The unequal mass for the $(c\bar{s})$ final state is a particularly
interesting laboratory for studying the diffractive factorization
breaking. A self-tagging property of charm
jets gives better access to various diffractive structure
functions, for instance, to $F_{3}^{D(3)}$.

The subject of this paper is the derivation of the above stated
features of CC DDIS and its distinction from the EM one.
 For convenience we focus on the process  
$e^{+}p\rightarrow \bar{\nu} p'X$ already observed by ZEUS \cite{CCZEUS}. 
The discussion and results being easily translated to the 
$e^{-} p\rightarrow {\nu} p'X$ process. 
In diffractive CC $e^{+}p$ 
scattering the experimentally 
measured quantity is the five-fold differential cross section
$d\sigma^{(5)}(ep\rightarrow \nu p'X)/
dQ^{2}dxdM^{2}dp_{\perp}^{2}d\phi$. Here $X$ is the diffractive
state of mass $M$, $p'$ is the secondary proton with the transverse
momentum $\vec{p}_{\perp}$, $t=-\vec{p}_{\perp}^{2}$, $\phi$ is the 
angle between the $(e,e')$ and $(p,p')$ planes, $Q^{2}=-q^2$ is the 
virtuality of the $W^{+}$ boson, $x, y, x_{\Pom}$ and 
$\beta=x/x_{\Pom}$ are the standard diffractive DIS variables.

The underlying subprocess is diffraction excitation of the $W^+$
boson, $W^+ p\rightarrow p'X$. In the parity conserving EM DIS, the
exchanged photon have either longitudinal (scalar), 
$s={1 \over Q} (q_{+}n_{+} - q_{-} n_{-})$ or
transverse, in the $(e,e')$ plane, polarization $t_{\mu}$ (here
$n_{\pm}$ are the usual lightcone vectors, $n_{+}^{2}=n_{-}^{2}=0$,
$n_{+}n_{-}=1, q=q_{+}n_{+}+q_{-}n_{-}$ and $q.s = q.t = 0$). In the
parity-nonconserving CC DIS, the exchanged $W^{+}$ bosons have also
the out-of-plane linear polarization $w_{\mu}=\epsilon_{\mu\nu\rho\sigma}
t_{\nu}n_{\rho}^{+}n_{\sigma}^{-}$. We introduce also the usual
transverse metric tensor 
$\delta_{\mu \nu}^{\perp} =\delta_{\mu\nu}
+n_{\mu}^{-}n_{\nu}^{+} + n_{\mu}^{+}n_{\nu}^{-}= - t_{\mu} t_{\nu}
-w_{\mu} w_{\nu}$. Then, the polarization state of the $W^{+}$
is described by the leptonic tensor

\bea
{L_{\mu \nu}}= {2 Q^2 \over  y^2} \left[ -{1\over 2}
\delta^{\perp}_{\mu\nu}(1-y+{1\over 2}y^2)
+ {1 \over 2}(1-y) (t_\mu t_\nu - w_\mu w_\nu) +
(1-y) s_\mu s_\nu \;\; \right . \nonumber\\ 
\left . + {1 \over 2} (1-{1\over 2} y) \sqrt{ 1- y}(s_\mu t_\nu+ s_\nu t_\mu ) +
 {i \over 2} y (1- {1 \over 2} y) (w_\mu t_\nu - w_\nu t_\mu) +
 {i \over 2} y \sqrt{1-y} (w_\mu s_\nu - s_\mu w_\nu) \right ]
\label{eq:Lep}
\eea
which, upon contraction with the hadronic tensor leads to
6 different components for 

\noindent $d\sigma_{i}^{(3)} (W^{+}p\rightarrow p'X)/dM^{2}dt d\phi$
 \ \ labeled by $i=T,L,TT',LT,3$ and $LT(3)$ :

\bea
y{d\sigma^{(5)}(e^{+}p\rightarrow \bar{\nu} p'X)\over
dQ^{2}dydM^{2}dp_{\perp}^{2}d\phi} =
{G_{F} M_W^{2} Q^{2}  \over 4 \sqrt{2}\pi^2 (M_{W}^{2}+Q^2)^2}
\left\{(1-y+{1\over 2}y^{2})\cdot d\sigma_{T}^{D(3)}
-y(1-{1\over 2}y)\cdot d\sigma_{3}^{D(3)}\right. \nonumber\\
+(1-y)\cdot d\sigma_{L}^{D(3)}+
(1-y)\cos 2\phi \cdot d\sigma_{TT'}^{D(3)}\nonumber\\
\left.
+(1-{1\over 2}y)\sqrt{1-y}\cos\phi
\cdot d\sigma_{LT}^{D(3)}
 - y\sqrt{1-y}\sin\phi \cdot d\sigma_{LT{(3)}}^{D(3)}\right\}
/dM^{2}dp_{\perp}^{2}d\phi\, ,
\label{eq:SigDIS}
\eea
where $G_{F}$ is the Fermi coupling, $M_{W}$ is the mass of the 
$W$-boson.
Each and every $d\sigma_{i}^{(3)}(W^{+}p\rightarrow p'X)$ defines a set of
dimensionless diffractive structure functions $F_{i}^{D(4)}$ : 

\beq
(Q^{2}+M^{2})\ {d\sigma_{i}^{(3)}(W^{+}p\rightarrow p'X)\over
dM^{2}dp_{\perp}^{2}}= 
{\pi G_{F} M_{W}^{2} Q^2 \over \sqrt{2}(Q^{2} + M^2_W)^2}\cdot 
{\sigma_{tot}^{pp}\over 16\pi}\cdot
F_{i}^{D(4)}(p_{\perp}^{2},
x_{\Pom},\beta,Q^{2})\, ,
\label{eq:Fdiff4}
\eeq
It is also useful to introduce the t-integrated SF's 
\footnote{Our definition (\ref{eq:Fdiff3}) of $F_{i}^{D(3)}$ 
differs from the ZEUS/H1 \cite{ZEUSF2Pom} by the factor $x_{\Pom}$, 
so that $F_{i}^{D(3)}$  does not blow up at $x_{\Pom} \to 0$} 

\beq
F_{i}^{D(3)}(x_{\Pom},\beta,Q^{2})=
{\sigma_{tot}^{pp}\over 16\pi }\int  dp_{\perp}^{2}
F_{i}^{D(4)}(p_{\perp}^{2},x_{\Pom},\beta,Q^{2})\, .
\label{eq:Fdiff3}
\eeq
The diffractive SF's $F_{T}^{D(3)}, F_{L}^{D(3)}$ and
$F_{3}^{D(3)}$ are counterparts of the familar $F_{T}=F_{2}-F_{L},
F_{L}$ and $F_{3}$ for inclusive DIS of neutrinos,
$F_{3}^{D(3)}$ and $F_{LT(3)}^{D(3)}$, are C- and P-odd and vanish
in EM scattering. The discussion of the azimuthal
angle-dependent terms $TT',LT$ and $LT(3)$ goes beyond the scope
of this letter, in which we focus on 
$F_{T}^{D(3)}, F_{L}^{D(3)}$ and $F_{3}^{D(3)}$.

Up to now only relatively large $x \sim 10^{-2}$ are easily 
accessible in CC DIS \cite{CCZEUS,CCDDexp}. As for selecting 
diffractive events, one requires $x_{\Pom} <$(0.05-0.1), the 
kinematical relation
$\beta=x/x_{\Pom}$ implies that the experimentally observed CC
diffractive DIS will proceed at rather large $\beta$,
dominated by the partonic subprocess $W^{+}p\rightarrow (u\bar{d}) p',
(c\bar{s}) p'$. The relevant pQCD diagrams are shown in Fig.~1. In 
the following, we 
focus on the $c\bar{s}$ excitation, analogous considerations 
apply to $u \bar{d}$. $z$ and $(1-z)$ are the fractions 
of the (light--cone) 
momentum of the $W^{+}$ carried by the charmed quark and strange
antiquark respectively, $\vec{k}$ is the relative transverse momentum in
the $q\bar{q}$ pair (Fig.~1). The invariant mass of the dijet final states 
equals

\beq
M^2={\frac{k^2 + \mu^2}{z(1-z)}}  \, ,
\label{eq:M2}
\eeq
where $\mu^2=(1-z)m_c^2 + zm_s^2$. $m_c, (m_s)$ being the 
charmed (strange) quark mass. 
All SF's are calculable in terms of the same  quark helicity
changing and conserving amplitudes $\vec{\Phi}_{1}$ and $\Phi_{2}$
introduced in \cite{NZ92,NZsplit}. Combining the formalism of \cite{NZsplit}
with the treatment of charm leptoproduction in \cite{F2cs}, we have 
obtained (integration over the azimuthal orientation of $\vec{k}$ is
understood, $\alpha_{cc}={G_F M^2_W \over 2 \pi \sqrt{2}}$)

\bea
\frac{d\sigma^D_{L,T}}{dz dk^2 dt}\bigg |_{t=0} =
\frac{\pi^2 \alpha_{cc}}{3} 
\alpha_{S}^{2}(\bar{Q}^{2})
\left[ 
A_{L,T}\, (z,m_s,m_c)\vec{\Phi}_{1}^{2}  +
B_{L,T}\, (z,m_s,m_c)\Phi_{2}^{2}
\right] \, ,
\label{eq:DSigLT}
\eea
where
\bea
A_T(z) &=& \left[ 1 -2z(1-z) \right] \, ,
\label{eq:AT}
\\
B_T(z,m_{s},m_{c}) &=& \left[ m_{c}^2 - 2 z(1-z) m_{c}^2 - z^2 \Delta^2 \right]\, ,
\label{eq:BT}
\\
A_3(z) &=& \left[2z - 1  \right] \, ,
\label{eq:A3}
\\
B_3(z,m_{s},m_{c}) &=& \left[z^2 m_{s}^2- (1-z)^{2}m_{c}^{2}\right]\, ,
\label{eq:B3}
\\
A_L(z,m_{s},m_{c}) &=&{1\over Q^{2}}(m_{s}^{2}+m_{c}^{2}) \, ,
\label{eq:AL}
\\
B_L(z,m_{s},m_{c}) &=& 4Q^2 z^2 (1-z)^2
+4z(1-z)\mu^2 +{1\over  Q^{2}}[\mu^4+m_{c}^{2}m_{s}^{2}]
\label{eq:BL}
\eea
with $\Delta^2 = m_c^2 -m_s^2$. The amplitudes $\vec{\Phi}_{1}$ and $\Phi_{2}$ were 
derived in \cite{NZ92,NZsplit} and, to a logarithmic accuracy,   

\bea
\vec{\Phi}_{1} &\approx& 2\vec{k} (1-\beta)^2 
\frac{[(k^2+\mu^2)\beta + (1-\beta)\mu^2]}{(k^2 + \mu^2)^3} 
\int {d\tau \over \tau}W_{1}(\omega,\tau)G(x_{\Pom},\tau \bar{Q}^{2})
\label{eq:phi1}
\\
\Phi_{2} &\approx& (1-\beta)^2 
\frac{[(k^2+\mu^2)(1-2 \beta) - 2\beta \mu^2]}{(k^2 + \mu^2)^3}
\int {d\tau \over \tau}W_{2}(\omega,\tau) G(x_{\Pom},\tau \bar{Q}^{2}) 
\label{eq:phi2}
\eea
where $G(x,Q^{2})=xg(x, Q^2)$ is the gluon distribution in the proton and  
$\varepsilon^{2}=z (1-z) Q^2 + \mu^2$. As in the EM case 
\cite{GNZcharm,GNZlong,Bartels,Levin} the relevant 
effective pQCD factorization scale is found to be 
 
\beq
\bar{Q}^{2} = \varepsilon^{2}+k^{2}= \frac{k^2 + \mu^2}{(1-\beta)} \,
\label{eq:Q2scale}
\eeq
and has already been used in (\ref{eq:DSigLT}) as the argument of 
strong coupling $\alpha_S$.

Here the weight functions 
$W_{i}(\omega = k^2/\varepsilon^2 ,\tau=\kappa^2/\bar{Q}^2)$ have a narrow 
peak at $\tau \approx 1 $ with the unit area under the peak, which gives 
the Leading Log$Q^{2}$ result \cite{NZsplit,GNZcharm, HT4}

\beq
\int {d\tau \over \tau}W_{i}(\omega,\tau)G(x_{\Pom},\tau \bar{Q}^{2})
\approx G(x_{\Pom},\bar{Q}^{2})\, ,
\label{eq:Geff}
\eeq
valid for sufficiently large values of $\omega$, which
is equivalent to sufficiently large $\beta \gsim 0.1$ of the
interest in the present study.

At variance with the equal mass EM case, where 
$\bar{Q}^{2}=(k^{2}+m^{2})/(1-\beta)$, 
now the factorization scale depends on $z$ and then one expects 
different cross sections whether the charmed
quark is produced in the forward (F) or the backward (B) hemisphere,
with respect to the $W$ momentum, in the rest frame of the diffractive
state $X$. The two configurations differ by the value of the light-cone
variable $z_{F,B}={1\over 2}(1+\delta)
\left[1\pm\sqrt{ 1 -4 {k^2 + m_c^2 \over M^2 (1 + \delta)^{2}}}\right]$,
where 
$\delta= {\Delta^2 \over Q^2}\; {\beta \over (1-\beta)}$.
The pQCD scale is  perturbatively large for large
$\beta$ even for light flavours, and for the charm component of the
diffractive SF it is large for all $\beta$, see below.

\noindent 
To evaluate the light quark component of the diffractive SF 
at not really large $\beta$, one needs a model for the small-$Q^2$ 
behaviour of the gluon structure function $G(x,Q^{2})$: in the following 
we will use the same form used in Ref. \cite{HT4}, which at large $Q^2$ 
coincide with the GRV NLO parameterization \cite{GRVNLO}. Furthermore we 
take $m_c=1.5$ GeV, $m_s=0.3$ GeV and $m_{u,d}= 150$ MeV. Variations of the 
charm mass by 10$\%$ have a small effect on the predicted SF, apart from 
shifting the threshold $\beta_{c}=Q^{2}/[Q^{2}+(m_{c}+m_{s})^{2}]$ (see below).

In the evaluation of $F_{i}^{D(3)}$ one needs to know the
$p_{\perp}^{2}$ dependence of the diffractive cross section,
which is usually parameterized as $d\sigma/dp_{\perp}^{2}
\propto \exp(-B_{D}p_{\perp}^{2})$. As it was shown in
\cite{NPZslope,NZDIS97}, one can use $B_{D} \sim 6$ GeV$^{-2}$ 
for heavy flavour excitation and for 
the perturbative transverse higher twist and logitudinal contributions 
while for light flavour contribution 
the diffraction slope $B_{D}$ exhibits, at not so large $\beta$, 
a slight $\beta$-dependence, but for the purposes of this present 
exploratory study we shall simply take $B_{D}(ud) \sim 9$ GeV$^{-2}$.

Many authors treats diffractive DIS as DIS off pomerons
in the proton, assuming implicitly and explicitly the diffractive
factorization. The latter is not supported by QCD studies
\cite{GNZ95,GNZcharm}, and the present study of charm excitation
in CC diffractive DIS offers more
evidence to this effect. Still it is not confusing, we shall 
speak of the perturbative intrinsic partons in the pomeron.

Separation of the pQCD subprocess of $W^{+}\rightarrow c\bar{s}$ into the
excitation of charm on the perturbative intrinsic strangeness in the pomeron and
excitation of (anti)strangeness on the intrinsic (anti)charm is not
unambiguous and must be taken with the grain of salt. In the naive
parton model, in the former process charmed quark will carry the
whole momentum of the $W^{+}$ and be produced with $z\approx1$.
In contrast, in the latter process, it is the strange antiquark
which carries the whole momentum of $W^{+}$ and charmed quark is
produced with $z\approx 0$, which suggests $z>{1\over 2}$ and
$z<{1\over 2}$ as a compromise boundary between the two partonic
subprocesses. However, the full fledged pQCD calculation
leads to broad $z$ distributions (for a related discussion of definition
of the strangeness and charm density in $\nu N,\bar{\nu}N$ inclusive DIS
see \cite{F2cs}). As a purely operational definition, we stick to a
parton model decomposition 
$F_{T}^{D(3)} (c \bar{s}) = F_{T(s)}^{D(3)}+ F_{T(\bar{c})}^{D(3)}$ 
and 
$F_{3}^{D(3)} (c \bar{s}) = F_{T(s)}^{D(3)}- F_{T(\bar{c})}^{D(3)}$, 
which is a basis for the results shown in
Fig.~2. With this definition, excitation of the
charmed quark off the intrinsic strangeness, $F_{T(s)}^{D(3)}$,
comes from terms $\propto z^2$ in (\ref{eq:AT}, \ref{eq:BT})
and (\ref{eq:A3}, \ref{eq:B3}). It is dominated by the forward
production of charm w.r.t. the momentum of $W^{+}$ in the
rest frame of the diffractive system $X$, but receives certain
contribution also from $z< {1\over 2}$. Similarly, $F_{T(\bar{c})}^{D(3)}$
some from terms $\propto (1-z)^2$, is dominated by the forward
production of strangeness (the backward production of charm), but
still receives certain contribution from the forward charm production.

\bigskip 

All the considerations of Ref. \cite{GNZlong, HT4} for the longitudinal 
and transverse diffractive SF in electroproduction are fully
applicable to the CC case at $Q^{2}\gg m_{c}^{2}$. 
We consider first the backward charm, $z\ll 1$, for which Eq.s 
(\ref{eq:M2}, 
\ref{eq:Q2scale}) give  $z \approx (k^2 + \mu^2)/M^2$ and 
$\bar{Q}^{2} \approx (k^{2}+m_{c}^{2})/(1-\beta)$.
Expanding the brackets of Eqs.(\ref{eq:phi1}, \ref{eq:phi2}), 
in Eq.(\ref{eq:DSigLT}) and using the approximation (\ref{eq:Geff}) 
the $k^2$-integration in (\ref{eq:DSigLT}) gives dominant contributions 
to the transverse SF coming from the low-$k^2$ region but without entering 
deeply in the nonperturbative region for the heavy quark production. For 
$M^2 \gg m_c^2$ one finds for the low scales dominated contribution 
(including the Leading Twist  and the first Higher Twist) :

\begin{eqnarray}
F_{T(\bar{c}) }^{D(4)} &\approx&   \frac{4 \pi}{3\sigma^{pp}_{tot}}\, 
\frac{\beta (1-\beta)^{2}}{6 m_c^2 (1+\delta)} 
\left\{ (3 + 4\beta + 8\beta^2) + 
{m^2_c \over Q^2} \frac{4 \beta}{1-\beta}   \right. 
\\ \nonumber 
& & \left. 
 \times  \left[ {5 \over 4} {\Delta^2 \over m^2_c}(1+ 8\beta^2) 
- (1-2\beta+4\beta^2)\right] \right\}
\left[ \alpha_S (\bar{Q}^2_{L}) 
G(x_{\Pom},\bar{Q}_{L}^{2} \simeq {m_c^2 \over (1-\beta)}) \right]^2 
\label{eq:FTL}
\end{eqnarray}

As in the EM case the large $k^2$  ($k^2 \sim M^2/4$) dominated 
contributions come from the second term in (\ref{eq:AT}). They are 
calculable in pQCD and the twist expansion starts with twist-4:  

\beq
F_{T(\bar{c}) }^{D(4)} = - \frac{16 \pi}{\sigma^{pp}_{tot}}\;
{2 \over 3} \frac{\beta^2 (1-\beta)}{Q^2 (1+\delta)} 
\left( \beta^2 + 2\; {\Delta^2 \over Q^2} \;  
\frac{\beta^2 (2\beta-1)}{(1-\beta)} \right)   
\left[\alpha_S (\bar{Q}^2_{H}) 
G(x_{\Pom},\bar{Q}_{H}^{2}\simeq {Q^2\over 4 \beta})\right]^2 
\label{eq:FTH}
\eeq

In (17, 18) emerge additional higher-twist corrections 
$\propto (\Delta^2/Q^2)^n$, in a first approach we restrict ourselves to the 
leading twist and its first higher twist corrections.

\noindent 
Thus, the higher twist corrections to $F^D_T$ receive contributions both from the 
low-scale region and the large scale $\bar{Q}^2_H$. The first term in 
Eq.(\ref{eq:FTH}) is substantially the same which has been discussed in 
Ref. \cite{HT4} for EM current and it remains relevant even at relatively large 
value of $Q^2$ as the $1/Q^2$ factor is partially compensated by the growth of 
$G(x_{\Pom}, \bar{Q}^2_H)$.

For what concerns the longitudinal cross section, the most important
contribution comes from the term $z^{2}(1-z)^{2}Q^{2}$ in the $B_{L}$
expansion (\ref{eq:BL}), which is identical to that
in the EM case. The $k^{2}$ integrated cross section is completely 
dominated by the short-distance contribution from high-$k^{2}$ jets,
$k^{2} \sim {1\over 4}M^{2}$. Upon the $k^{2}$ integration, to a
logarithmic accuracy, we find the twist expansion of the 
longitudinal SF for the terms dominated by the large scale:

\beq
 F_{L(\bar{c})}^{D(4)} = \frac{16 \pi}{\sigma^{pp}_{tot}}\; 
\frac{\beta^{3}\;(2\beta -1)}{3\ Q^{2}(1+\delta)}
\left((2\beta-1) + 
{\Delta^2\over Q^2}\frac{\beta(5 - 6\beta)}{(1-\beta)}\right)
\left[\alpha_S(\bar{Q}^2_H) G(x_{\Pom},\bar{Q}^{2}_H) \right]^2
\label{eq:FLH}
\eeq

As the pQCD scale $\bar{Q}_{H}^{2}$ does not depend on 
flavours we predict a restoration of the flavour symmetry and equal 
$c\bar{s}$ and $u\bar{d}$ when twist-6 is negligible.
Such an equal contribution of light and heavy flavours into the higher 
twist is unprecedented in the standard inclusive DIS. Again the scaling 
violation factor $G^{2}(x_{\Pom},\bar{Q^{2}})$, in (\ref{eq:FLH}),
largely compensates the higher twist factor ${1\over Q^{2}}$ 
and the longitudinal SF remains large, and takes over $F_{T}^{D}$,
in a broad range of $Q^{2}$ of practical interest, see Fig.~2.

In (\ref{eq:FLH}), the leading twist-4 term is the same as for NC diffractive DIS.
However, in the CC diffractive DIS, because non-conservation of weak 
current,  extra higher twist contributions to $F_{L}^{D(3)}$ come from 
the expansion of $B_{L}$ (always substantially dominated by the 
perturbative region). 

Further terms (both twist-4 and higher), come from 
the term $\propto A_{L}$ in (\ref{eq:DSigLT}). They 
receive large contributions from the low $k^2$ region.
In particular they assume a strong relevance for the charm--strange component
where terms $\propto m_c^2/Q^2$ appear. 
We find for the $A_L$ contribution: 

\beq
F^{D(4)}_{L (\bar{c})} [A_L] \approx \frac{4 \pi}{9 \; \sigma^{pp}_{tot}} \;
\frac{(m_c^2 + m_s^2)}{m_c^2} \; 
\frac{\beta (1-\beta)^2}{Q^2 (1-\delta)} (1+2\beta+3\beta^2)
\left[ \alpha_S(\bar{Q}^2_L) G(x_{\Pom}, \bar{Q}^2_L) \right]^2
\label{eq:FLAL}
\eeq


Whereas this comtribution is low scale dominated it is comparable 
to the leading twist-4 in the small $Q^2$ region. Due to the symmetry 
$z, (1-z)$ of Eqs.(\ref{eq:M2},
\ref{eq:AT} - \ref{eq:BL}), one finds similar results subject to the 
replacement $m_{c} \rightarrow m_{s}$ for the forward production of 
charm ($F^D_{(s)}$) at $1-z \lsim m_{s}^{2}/m_{c}^{2}$. The overall 
dependence of $F_L^{D(3)}(cs)$ on $Q^2$ and its decomposition in the 
$A_L$ and $B_L$ components are shown in Fig.~3.


\bigskip

It is interesting to notice that the pQCD scales  $\bar{Q}^2{(c)}$ and
$\bar{Q}^2{(s)}$ are different, both explicitly depend on $\beta$, 
and the $x_{\Pom}$ and $\beta$ dependences of $F_{T(i)}^{D(3)}$ are 
inextricably entangled. This gives another example where
the Ingelman-Schlein factorization hypothesis, 
$F_{2}^{D(3)}(x_{\Pom},\beta,Q^{2}) = 
f_{\Pom}(x_{\Pom}) F_{2 \Pom}(\beta,Q^{2})$, with process
independent flux of pomerons in the proton $f_{\Pom}(x_{\Pom})$ and the 
$x_{\Pom}$ independent pomeron SF $F_{2\Pom}(\beta,Q^{2})$, is not confirmed 
by pQCD calculation. The diffractive factorization breaking in CC diffractive 
DIS is especially severe, because for the same $c\bar{s}$ final state the pQCD 
factorization scale $\bar{Q}^{2}$ changes substantially from the forward to 
backward hemisphere: $\bar{Q}{(s)}^{2} \ll \bar{Q}{(c)}^{2}$. Although the 
perturbative intrinsic charm component $F_{T(\bar{c})}^{D(3)}$ is suppressed by
the mass of a heavy quark, it is still substantial and it is predicted to rise 
much steeper than the strange one as $x_{\Pom}\rightarrow 0$. Furthermore, 
$F_{T(\bar{c})}^{D(3)}$ is truly of perturbative origin at all $\beta$, while 
$F_{T(s)}^{D(3)}$ has a non-negligible dependence to small scales up to 
$\beta \gsim 0.7$.

As an illustration of the diffractive factorization breaking, in
Fig.~4  we show the effective exponent of the $x_{\Pom}$ dependence
\beq
n_{eff} = 1-{\partial \log F_{2}^{D(3)} \over \partial\log x_{\Pom}}
\label{eq:neff}
\eeq
evaluated for $x_{\Pom}=3\cdot 10^{-3}$. We show the $\beta$ dependence 
of $n_{eff}$ evaluated for $F_{2}^{D(3)}(u\bar{d} +c\bar{s})$ 
and $F_{2}^{D(3)} (u\bar{d})$ for $\beta > 0.2$. At smaller $\beta$ one 
expects a further increasing due to the triple pomeron component, 
see \cite{GNZA3Pom}.

Evidently, at fixed $\beta$, the $c\bar{s}$ excitation is possible
only for sufficiently large $Q^{2}$ such that $\beta_{c} > \beta$.
For this reason, diffractive SF's exhibit strong
threshold effects shown in Fig.~5, which are much stronger than
in the NC case studied in \cite{GNZcharm, HT4}. Notice, that $F_{3}^{D(3)}$
vanishes below the $c\bar{s}$ threshold.


Naively, one would expect $F_{3}^{D(3)}=0$ for a quark-antiquark
symmetric target  as the Pomeron is. Indeed, because $A_{3}(z)$, and
for equal mass case, $B_{3}(z)$ too, are antisymmetric
about $z={1\over 2}$, the contribution from
$u\bar{d}$ excitation to $F_{3}^{D(3)}$  vanishes upon the integration
over the $u$-jet production angles.
On the other hand, in the $c\bar{s}$
excitation there is a strong forward-backward asymmetry  and
$F_{3}^{D(3)} (c \bar{s}) \neq 0$.
Our predictions for $F_{3}^{D(3)}$ are shown in Fig.~2.

Finally, some comments on the so-called triple-pomeron region, of $\beta \ll 1$,
are in order. Here diffraction proceeds via excitation
of the soft gluon-containing $q\bar{q}g$ and higher Fock states of
the photon. As it has been discussed to great detail in \cite{NZ94,GNZ95},
at $\beta \ll 1$ and only at $\beta \ll 1$, and with certain reservations,
one can apply the standard parton model treatment to diffractive DIS.
For instance, the conventional fusion of virtual photons with the gluon
from the two-gluon valence state of the pomeron becomes the driving term
of diffractive DIS.

In this case the results for the diffractive SF of light
quarks coincide (once the opportune couplings of weak interaction are
substituted to the EM ones) with those presented in Ref. \cite{GNZ95}.
For the charm--strange component it must be considered that
now the charm quark  is always produced together with a strange one,
this leads to a threshold ($Q^2_{cs}= 4 GeV^2$), which is intermediate
between the strange ($Q^2_{ss}= 1 GeV^2 $ ) and the charm
($Q^2_{cc}=10 GeV^2$) electromagnetic DIS thresholds in analogy to our
discussion concerning the inclusive DIS \cite{F2cs,F2charm}; using the 
the notation of \cite{GNZ95} we find $A_{cs}=0.08$.

\bigskip
\bigskip

\noindent
{\large \bf  Summary and Conclusions.}

We have presented the calculation of diffractive structure functions
in the QCD color-dipole scheme for charged
current DIS and carried on a comparison of diffraction in charged current and
electromagnetic DIS. Both charged current and electromagnetic
diffraction share the property of diffractive factorization breaking.
For instance, we find different $x_{\Pom}$ dependences of the intrinsic
$u,d$, strangeness and charm composition of the pomeron. Futhermore, we predict a 
different $x_{\Pom}$ dependence even for the production of charm quark in the 
forward and backward direction.

Compared to the EM case, other new features of CC diffraction are the emergence of
substantial $F_{3}^{D(3)}$,  and the large higher twist contributions
to the longitudinal structure function because of the non-conservation
of weak currents. These predictions will be tested with the accumulation
of more data on CC diffraction at HERA and will permit further test of the color 
dipole picture of DDIS.



\newpage

\section*{Figures}

\begin{figure}[htbp]
  \begin{center}
    \mbox{\epsfig{file=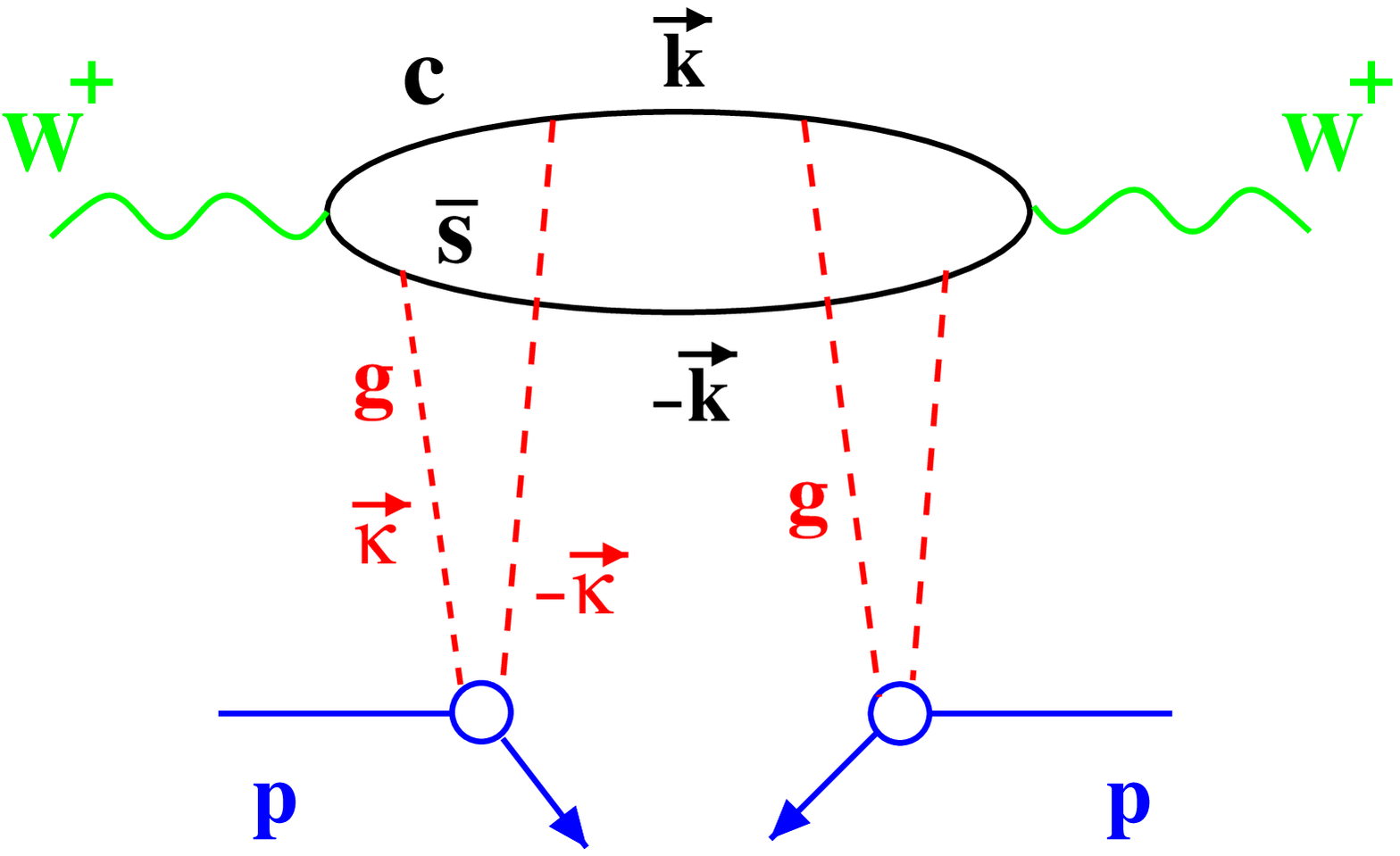,height=8cm,width=13cm,clip=}}
   \caption{The pQCD Feynman diagrams for diffraction exitation of 
            $c\bar{s}$ ($u \bar{d}$) states of the $W$.}
   \label{fig1}
\end{center}
\end{figure}

\newpage

\begin{figure}[htbp]
  \begin{center}
   \mbox{\epsfig{file=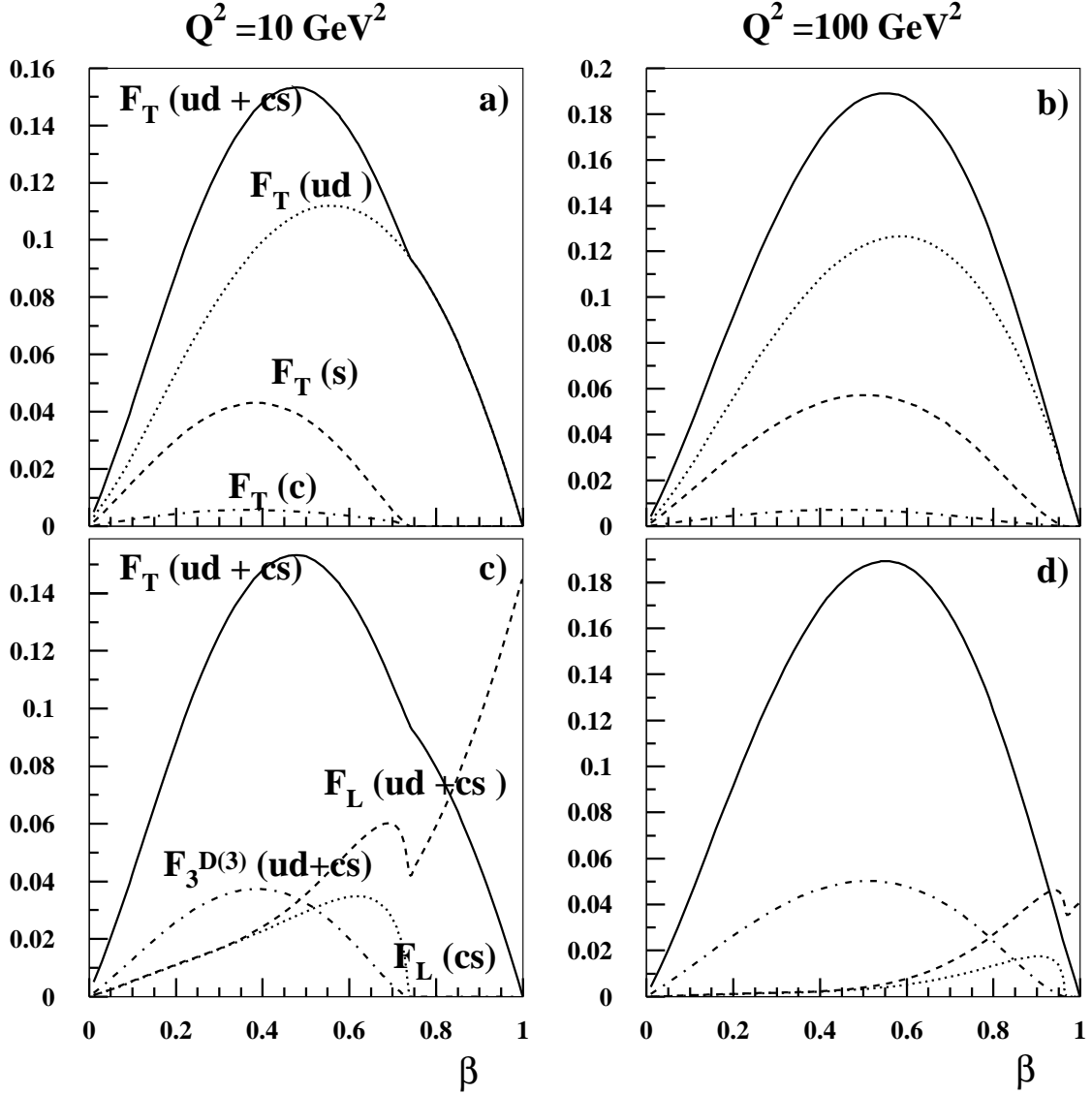,height=17cm,width=17cm,clip=}}
   \caption{The $\beta$ dependence for $x_{\Pom}=10^{-3}$ of 
           {\bf \ \ \ a)} $F_T (ud +sc)$ [solid], $F_T (ud)$ [dotted], 
           $F_T(s)$ [dashed], $F_T(c)$ [dot--dashed]  at $Q^2 =  10$ GeV$^2$ 
           {\bf  \ \  b)}  the same as above for  $Q^2 =100$ GeV$^2$
           {\bf \ \ c)} All flavours  $ F_T$ [solid], $F_3$ [dot--dashed], 
           $F_L$ [dashed] and $F_L(cs)$ [dotted]    at  $Q^2 = 10$ GeV$^2$
           {\bf \ \ d)}  the same as above for  $Q^2 = 100$ GeV$^2$}
   \label{fig2}
\end{center}
\end{figure}

\newpage

\begin{figure}[htbp]
  \begin{center}
    \mbox{\epsfig{file=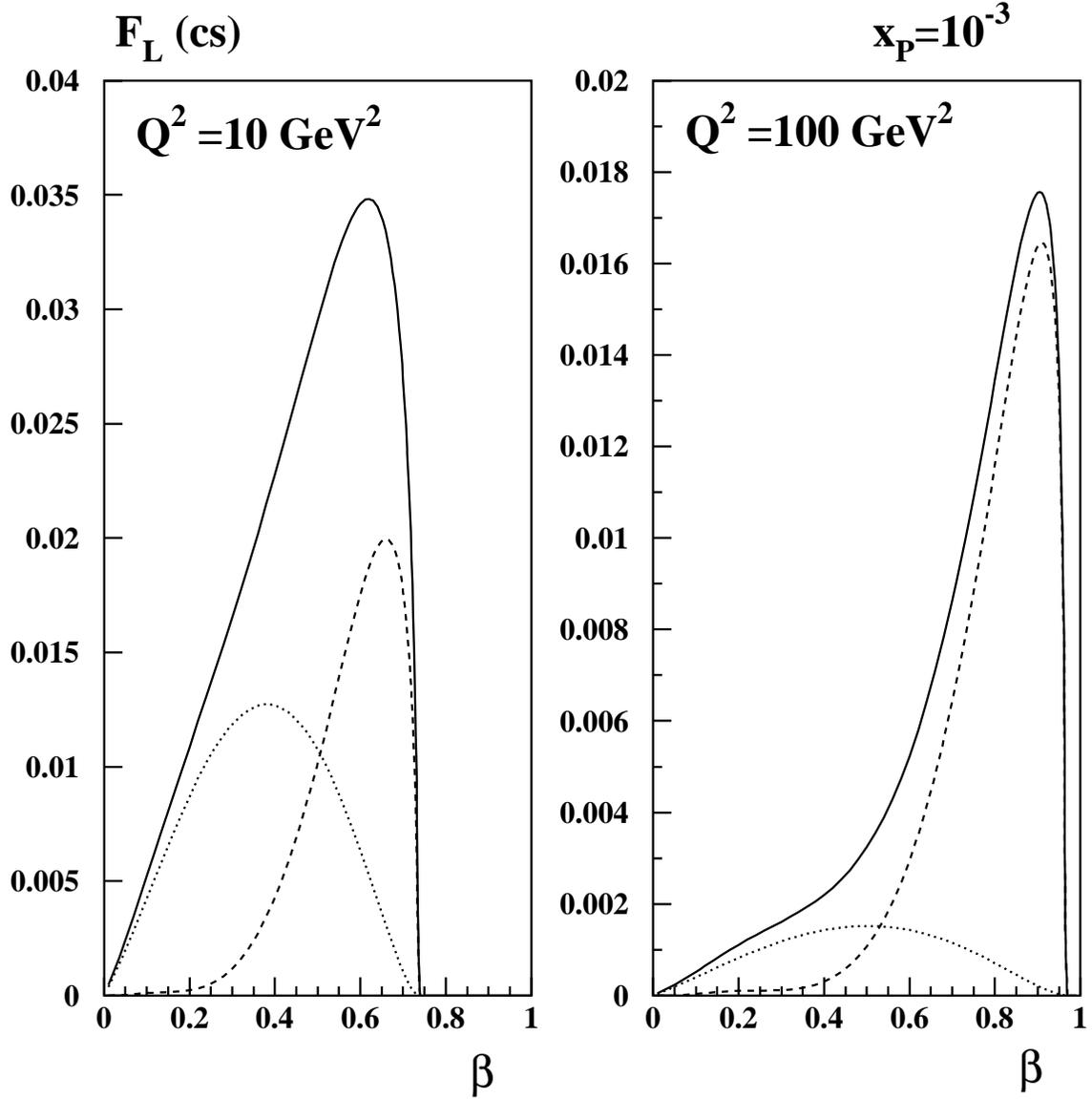,height=17cm,width=17cm,clip=}}
   \caption{$F_L^{D(3)}(cs)$ [solid] and $A_L$ Eq.(\ref{eq:FLAL}), 
            [dotted] and $B_L$ Eq.(\ref{eq:FLH}), [dot--dashed] 
            components of $F_L$  at $Q^2$=10,100  GeV$^2$.}
   \label{fig3}
\end{center}
\end{figure}

\newpage 

\begin{figure}[htbp]
  \begin{center}
    \mbox{\epsfig{file=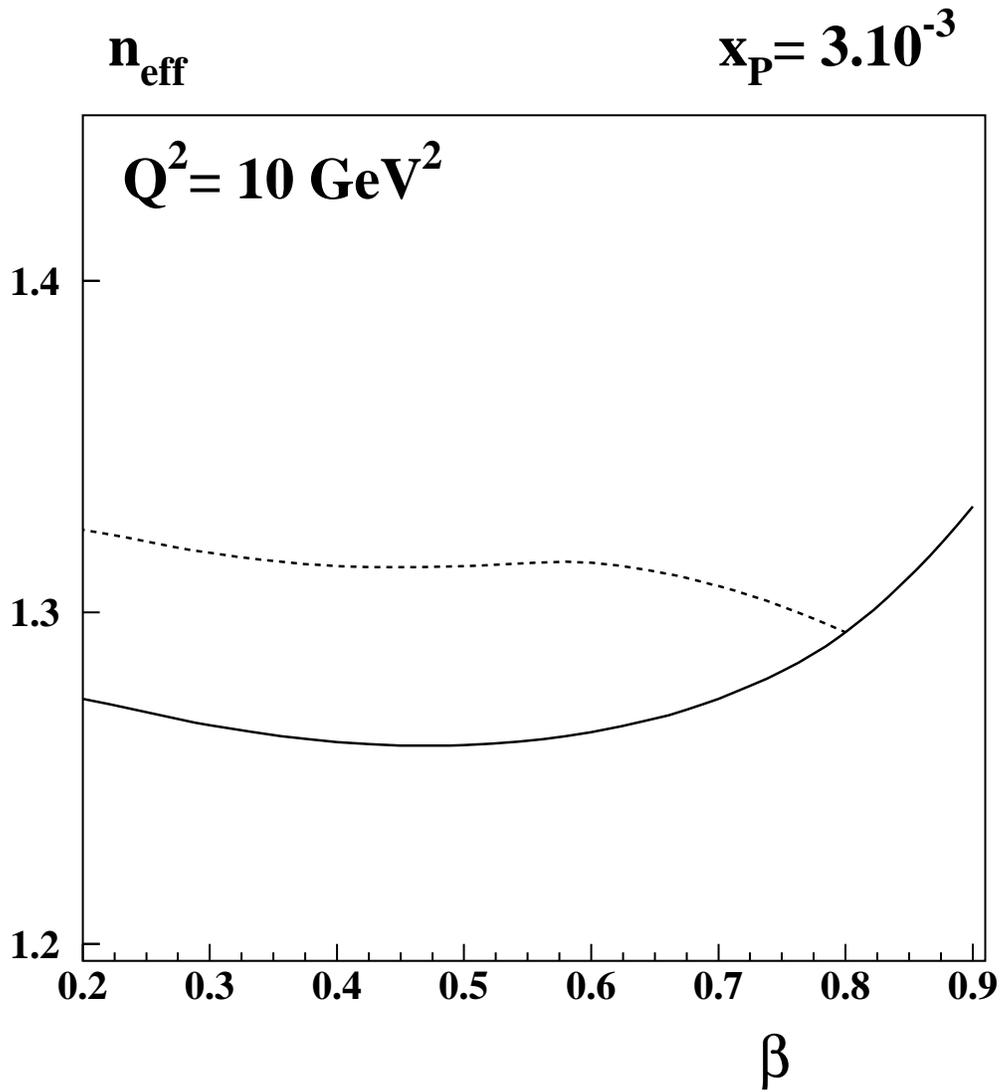,height=15cm,width=15cm,clip=}}
   \caption{The $\beta$ dependence of $ n_{eff}$  for $x_{\Pom}=0.003$ 
            and $Q^2=10$ GeV$^2$. Solid line : 
            (ud) component, dashed line: (ud+cs).}
   \label{fig4}
\end{center}
\end{figure}

\newpage

\begin{figure}[htbp]
  \begin{center}
    \mbox{\epsfig{file=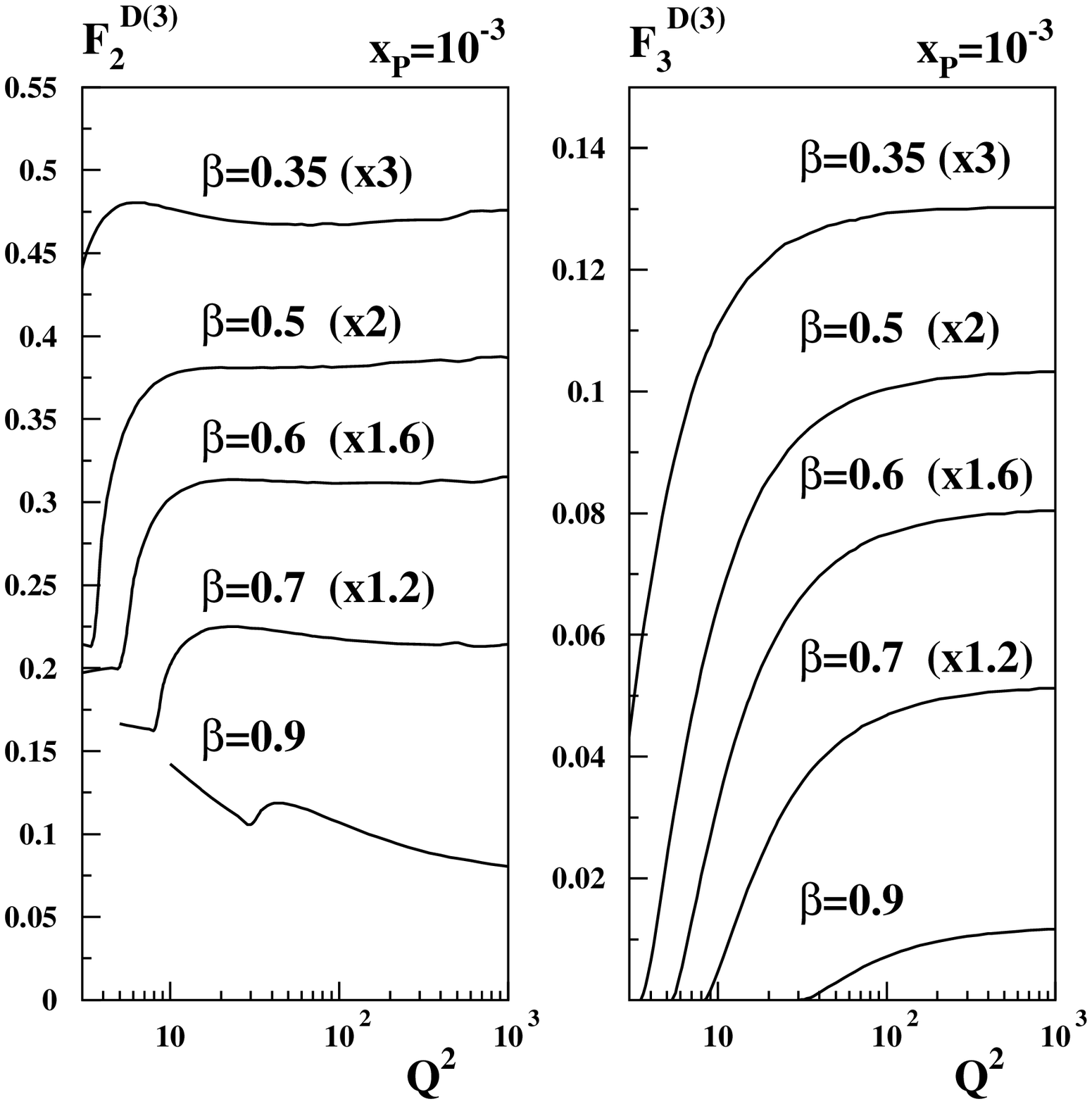,height=17cm,width=17cm,clip=}}
   \caption{Charm-strange threshold effect in the $Q^2$ dependence 
            of the diffractive SF $F_2^{D(3)}$ (first box) and on   
            $F_3^{D(3)}$ (second boxe).}
   \label{5}
\end{center}
\end{figure}


\begin{thebibliography}{299}
\bibitem{NZ92} 
N.N. Nikolaev  and B.G.~Zakharov, {\it Z. Phys.}
{\bf C53},  331 (1992).

\bibitem{NZsplit} 
N.N. Nikolaev and B.G.Zakharov,
{\sl Phys. Lett.} {\bf B332}, 177 (1994).

\bibitem{NZ94} 
N.N. Nikolaev  and B.G.Zakharov,  {\sl JETP}
{\bf 78}, 598 (1994);
{\sl Z. Phys.} {\bf C64} (1994) 631.

\bibitem{GNZ95} 
M. Genovese, N.N. Nikolaev  and B.G. Zakharov,
 {\sl JETP} {\bf 81} 625 (1995).

\bibitem{GNZA3Pom} 
M. Genovese, N.N. Nikolaev and B.G. Zakharov,
 {\sl JETP} {\bf 81},  633 (1995).

\bibitem{GNZcharm} 
M. Genovese M., N.N. Nikolaev  and B.G. Zakharov,
{\sl Phys. Lett.} {\bf B378}, 347 (1996).

\bibitem{GNZlong} 
M. Genovese, N.N. Nikolaev and B.G. Zakharov, {\sl Phys.Lett.}
{\bf B380},  213 (1996).

\bibitem{HT4}
M. Bertini, M. Genovese, N.N. Nikolaev, A.V. Pronyaev and B.G. Zakharov, 
{\sl Phys. Lett.} {\bf B422} (1998) 238.

\bibitem{Bartels} %
J. Bartels, J. Ellis, H. Kowalski and M. Wuesthoff {hep-ph/9803497},  
J. Bartels, H. Lotter and M. W\"usthoff, {\it Phys. Lett.} {\bf B379} (1996) 239, 
H. Lotter, {\it Phys. Lett.} {\bf B406} (1997) 171.

\bibitem{Levin}
E.M. Levin, A.D. Martin, M.G. Ryskin and T. Teubner, {\it Z. Phys.}
{\bf C74} (1997) 671. E.Gotsman, E.Levin and U.Maor, {\it Nucl. Phys.}
{\bf B493} (1997) 354.

\bibitem{NZDIS97}
Nikolaev N.N. and B.G.Zakharov, Phenomenology
of Diffractive DIS. Overview at DIS'97, Chicago, April 97, hep-ph/9706343;
M. Genovese, to be published in Proc. of DIS98, Bruxelles, April 98, hep-ph/9805504.

\bibitem{CCZEUS}
ZEUS Collab., M. Derrick et al., {Z. Phys.} {C72} (1996) 47.

\bibitem{ZEUSF2Pom} 
H1 Coll., C. Adloff et al. {\it Z. Phys.} {\bf C76} (1997) 613,  

ZEUS Coll., M. Derrick et al. {\sl Z. Phys.} {\bf C68} (1995) 569. 


\bibitem{CCDDexp}
J. Pliszka and A. F. Zarnecki, Proc. Workshop on Future Physics at
HERA (Hamburg 1996) ed. G. Ingelman {\it et al.}, p728. 


\bibitem{F2cs}V. Barone, M. Genovese,
N.N. Nikolaev, E. Predazzi and B.G. Zakharov, \PLB 328 (1994) 143.

\bibitem{Buchmuller}
W. Buchm\"uller, M.F. McDermott and A. Hebecker, hep-ph/9703314.

\bibitem{NPZslope}
N.N. Nikolaev, A. Pronyaev and B.G. Zakharov, paper in preparation.

\bibitem{BGNPZcc} 
V. Barone et al., {\sl Phys. Lett.} {\bf B292}, 181 (1992).
V. Barone, M. Genovese, N.N. Nikolaev, E. Predazzi and B.G. Zakharov,
{\sl Int. J. Mod. Phys. } {\bf A8} (1993) 2779.

\bibitem{GRVNLO} 
M. Gl\"uck, E. Reya and A. Vogt, {\sl Z. Phys.} {\bf C67}, 433 (1995).

\bibitem{F2charm}
V. Barone, M. Genovese, N.N. Nikolaev, E. Predazzi and B.G. Zakharov,
\ZPC 70 (1996) 83; Phys. Lett. {B304} (1993) 176, {B268} (1991) 279,
{B317} (1993) 433 ; V. Barone and M. Genovese, \PLB 379 (1996) 233.

\end{thebibliography}
\end{document}